\documentclass{article}
\usepackage{amsmath,graphicx,mlspconf}
\usepackage[numbers]{natbib}
\usepackage[bookmarks=false]{hyperref}
\usepackage{amssymb}
\usepackage{mathtools}
\usepackage{booktabs}

\copyrightnotice{978-1-7281-6338-3/21/\$31.00 {\copyright}2021 IEEE}

\toappear{2021 IEEE International Workshop on Machine Learning for Signal Processing, Oct.\ 25--28, 2021, Gold Coast, Australia}

\title{MLP Singer: \\ Towards Rapid Parallel Korean Singing Voice Synthesis}
\name{%
    Jaesung Tae$^{1}$
    \sthanks{Work performed while interning at Neosapience.}%
    \qquad Hyeongju Kim$^{2}$%
    \qquad Younggun Lee$^{2}$%
}
\address{%
    $^{1}$ Yale University, New Haven, CT, USA \\%
    $^{2}$ Neosapience, Inc., Seoul, Republic of Korea%
}

\begin{document}
% \ninept

\maketitle

\begin{abstract}
Recent developments in deep learning have significantly improved the quality of synthesized singing voice audio. However, prominent neural singing voice synthesis systems suffer from slow inference speed due to their autoregressive design. Inspired by MLP-Mixer, a novel architecture introduced in the vision literature for attention-free image classification, we propose MLP Singer, a parallel Korean singing voice synthesis system. To the best of our knowledge, this is the first work that uses an entirely MLP-based architecture for voice synthesis. Listening tests demonstrate that MLP Singer outperforms a larger autoregressive GAN-based system, both in terms of audio quality and synthesis speed. In particular, MLP Singer achieves a real-time factor of up to 200 and 3400 on CPUs and GPUs respectively, enabling order of magnitude faster generation on both environments.\footnote{Source code available at \url{https://github.com/neosapience/mlp-singer}.}
\end{abstract}
\section{Introduction}
\label{sec:intro}

Singing voice synthesis (SVS) is a generative task in which a model learns to map musical scores containing linguistic, temporal, and melodic information to acoustic features. While SVS bears similarity to text-to-speech (TTS), it is distinct in that both pitch and temporal information are already provided in the input. A typical TTS model has to learn prosodies as well as temporal alignments between linguistic and acoustic features. Contrastingly in SVS, it is apparent at every time step which syllable should be voiced at what pitch given a musical score. Therefore, the objective of an SVS model is to learn to project time-aligned phoneme and pitch representations to acoustic features while interpolating details not immediately present in the input, such as timbre, breath, and vocalization methods. Some systems relax this task by assuming continuous $f_0$ values as given \cite{chandna2019wgansing,blaauw2020sequencetosequence}; however, the problem we consider here is one in which only low-resolution pitch sequences are available, and the model has to learn how to dequantize that information to produce smooth acoustic features.

Traditional SVS systems rely on concatenation \cite{bonada2003sample} or statistical parametric methods such as hidden Markov models \cite{saino2006hmm}, which require large amounts of data or complex pipelines for high quality synthesis. On the other hand, deep neural networks have proven to be remarkably effective in speech synthesis due to their ability to approximate complex non-linear functions. Building on top of recent advancements in TTS architectures, modern SVS systems employ deep neural networks to autoregressively generate vocoder features from melody and lyrics. NPSS \cite{blaauw2017npss} uses causal dilated convolutions to produce WORLD vocoder features \cite{masanori2016world} conditioned on musical scores. Lee et al. \cite{lee2019adversarially} proposes an adversarial training scheme and phonetic enhancement masks to improve pronunciation, while Choi et al. proposes an autoregressive model trained on the boundary equilibrium GAN objective \cite{choi2020korean}.

While autoregressive sampling can generate smooth acoustic features, it has a number of drawbacks. Autoregressive models are prone to exposure bias caused by train-test discrepancy; during training, the model is teacher-forced ground-truth data, which are not available during inference \cite{schmidt2019generalization}. Moreover, generation can be time-consuming, especially for long sequences. Autoregressive sampling requires that each output be contingent on the output from the previous time step. This introduces temporal overhead since each input can only be processed after previous chunks have fully been produced. Blaauw et al. \cite{blaauw2020sequencetosequence} avoids these issues by using a feedforward transformer \cite{vaswani2017attention} to generate WORLD vocoder features from alignments produced by a duration predictor. However, the computational complexity of self-attention \cite{kitaev2020reformer} highlights the need for a light-weight baseline. 

We propose MLP Singer, a parallel Korean singing voice synthesis system exclusively composed of multi-layer perceptrons (MLPs). MLP Singer is based on MLP-Mixer \cite{tolstikhin2021mlpmixer}, an architecture introduced in the computer vision literature that demonstrates the potential of MLPs as a competitive replacement for transformers and convolutional neural networks. MLPs scale linearly with respect to input, and their simple structure benefits from hardware and software accelerations highly optimized for matrix multiplication. Through listening tests, we show that the proposed system serves as a strong baseline, achieving superior performance to a larger autoregressive conditional GAN-based system \cite{choi2020korean} both in terms of audio quality and inference latency. We also propose an overlapped batch segmentation method as a means of reducing discontinuous frame artifacts that typically affect non-autoregressive models.

\section{Architecture}
\label{sec:pagestyle}

\begin{figure*}[t]
  \centering
  \includegraphics[scale=0.7]{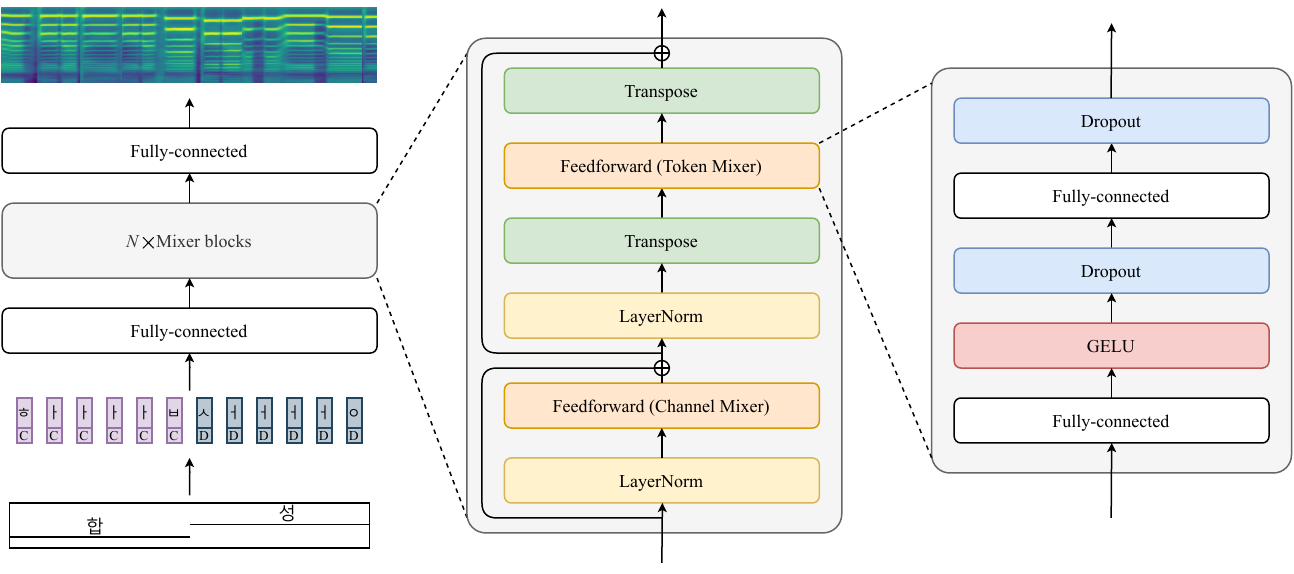}
  \caption{Overview of MLP Singer}
  \label{fig:overview}
\end{figure*}

An overview of MLP Singer is illustrated in Figure \ref{fig:overview}. The model receives lyrics text and a sequence of MIDI notes as input. Exploiting the fact that text and pitch sequences are time-aligned, we convert each sequence to embeddings and concatenate them to feed into the model. The first layer is a fully-connected layer that projects the concatenated embeddings to a latent space, from which subsequent Mixer blocks gradually apply transformations to map intermediate representations to acoustic features. Finally, the output is projected to the mel-spectrogram space through a fully-connected layer. In the sections that follow, we expound the components of the model in greater detail.

\subsection{Data Representation}

\subsubsection{Input Processing}

Given a total sequence length $T$, let $\mathbf{m} = (m_1, m_2, \dots, m_T)$ denote the sequence of MIDI events, and $\mathbf{c} = (c_1, c_2, \dots, c_T)$ denote the sequence of text as syllables. Each MIDI event $m_i$ can be decomposed into  $(p_i, s_i, e_i)$, where $p_i$ denotes the pitch of the $i$-th event as MIDI note numbers, $s_i$ the starting time of the event, and $e_i$ the ending time. $m_i$ and $c_i$ combined informs the duration of the $i$-th syllable and its pitch.

Because $T$ does not equal the number of mel-spectrogram frames $L$, $\mathbf{p} = (p_1, p_2, \dots, p_T)$ and $\mathbf{c}$ are elongated to match $L$. This expansion is possible because each mel-spectrogram frame can be attributed to a single index in $[1:T]$ by observing the starting times and ending times given from $\mathbf{m}$. The resulting sequences are $\mathbf{p^\prime}$ and $\mathbf{c^\prime}$, which represent spectrogram-aligned pitch and phoneme sequences, respectively. 

\subsubsection{Embedding}

In Korean, every syllable can be subdivided into at most three phonemes: an onset, nucleus, and an optional coda. Exploiting the fact that vowels consume the majority of extended vocalizations, Lee et al. \cite{lee2019adversarially} allocates one frame for the onset and coda, populating the rest with the nucleus. We follow a similar approach adopted by Choi et al. \cite{choi2020korean} and introduce consonant emphasis factor $k$ as a hyperparameter. The distribution scheme, then, is to allot $k$ frames for the onset and coda, and $n - 2k$ for the nucleus. Empirically, a $k$ value of 3 produced desirable results. Although this is not an exact alignment, we found that the model is capable of generating plausible acoustic features from this initial approximation given its sufficient receptive field.

The expanded sequence $\mathbf{c^\prime}$ is embedded through a lookup table into $\mathbf{E_c} \in \mathbb{R}^{L \times D_c}$, where $D_c$ refers to the dimensionality of the phoneme embedding. By the same token, $\mathbf{p^\prime}$ is transformed into $\mathbf{E_p}$ through an embedding layer. The phoneme and pitch embeddings are then concatenated to $\mathbf{E} = (\mathbf{E_c}, \mathbf{E_p}) \in \mathbb{R}^{L \times D}$, where $D = D_c + D_p$ and $D_p$ denotes the cardinality of the pitch embedding space. $\mathbf{E}$ is then passed through a single fully-connected layer to produce a latent representation that is a linear combination of the phoneme and pitch embeddings.

\subsection{Mixer Block}

We adopt the Mixer block from MLP-Mixer \cite{tolstikhin2021mlpmixer} with minimal modifications. A Mixer block comprises pre-layer normalization \cite{ba2016layer} and feedforward layers with residual skip connections. A single feedforward layer is defined by two affine transformations with an intermediate activation. The structure is schematically shown below.

\begin{equation}
    \texttt{feedforward}_i(\mathbf{X}) = \sigma(\mathbf{X}^\top \mathbf{W}_1^i) \mathbf{W}_2^i 
\end{equation}
$\sigma(\cdot)$ represents a non-linearity, in this case Gaussian Error Linear Units (GELU) \cite{hendrycks2020gaussian}. The dimensions of $\mathbf{W}_1^i$ and $\mathbf{W}_2^i$ are $\mathbb{R}^{D_i \times D_h}$ and $\mathbb{R}^{D_h \times D_i}$, respectively, where $D_h$ denotes the size of the hidden dimension and $D_i$ denotes the input dimension of the $i$-th feedforward layer. Dropout and bias terms are omitted for brevity. 

A Mixer block consists of two such feedforward layers, except that the second layer is applied on the transposed axis.

\begin{gather}
    \mathbf{Z} = \mathbf{X}_l + \texttt{feedforward}_1(\tilde{\mathbf{X}_l}) \label{eq:mixer1} \\
    \mathbf{X}_{l + 1} = \mathbf{Z} + \texttt{feedforward}_2(\tilde{\mathbf{Z}}^\top)^\top \label{eq:mixer2}
\end{gather}
where $\tilde{\cdot} \coloneqq \texttt{layernorm}(\cdot)$ and $\mathbf{X}_l$ denotes the input to the $l$-th Mixer block.

We follow the nomenclature in MLP-Mixer and refer to the feedforward layer in Equation \ref{eq:mixer1} as the Channel Mixer and \ref{eq:mixer2} as the Token Mixer. The Channel Mixer is applied to each embedding, whereas the Token Mixer is applied in the transposed axis across different time steps. In MLP-Mixer, the Token Mixer precedes the Channel Mixer; in our experiments, we found that exchanging the order of these components yielded some benefits. We hypothesize that this is due to the importance of sufficiently diffusing text and pitch information across embedding channels prior to passing through the first Token Mixer layer.

At the core of this architecture is the Token Mixer. Unlike convolutional neural networks, which gradually enlarge their receptive fields via dilations or deep stacking, the Token Mixer gains immediate access to the entire input chunk by accepting latent representations in their transposed form. The wide context window afforded to the model starting from the very first Mixer block helps MLP Singer generate latent features while being aware of neighboring frames within the input window. 

The resulting output from $N$ Mixer blocks is projected by a final fully-connected layer that sends $\mathbf{X}_{N+1} \in \mathbb{R}^{L \times D}$ to $\mathbb{R}^{L \times D_\text{mel}}$, where $D_\text{mel}$ denotes the mel-spectrogram frequency bin. The model is trained on L1 mel-spectrogram loss. 

\subsection{Overlapped Batch Segmentation}
\label{sec:batch}

One corollary of transposed matrix multiplication is the constraint that inputs be padded or truncated to length $L$. Since a typical song is much longer than $L$, in practice, generating audio for an entire song requires padding and reshaping the input into batches of segments that are $L$ in length. Batching inputs in this fashion is possible only because MLP Singer generates acoustic features in parallel; autoregressive models can only process one segment at a time, and thus require iterative loops. MLP Singer fully enjoys the benefit of hardware and software accelerations optimized for batch processing.

A potential downside of simple batching is that the model may produce audible artifacts around segment boundaries. Choi et al. \cite{choi2020korean} found that autoregressive models generated coherent spectrograms, whereas models that were not provided any previous context produced spectrograms with abrupt interruptions in harmonic patterns. Motivated by a similar concern, we also consider an improved scheme in which the input is batched into overlapping chunks, where the size of the overlap window is given by $w$. The first and last $w$ frames of each chunk are then dropped to remove redundant frames. We experiment with both batching schemes to explore the importance of overlapping windows in parallel generation. 

\section{Experiment}

\subsection{Dataset}

We used the Children's Song Dataset (CSD) \cite{choi2020children}, an open dataset composed of English and Korean children songs sung by a professional female singer. Each song is sung twice in two different keys. We used 50 Korean songs, which totals approximately two hours in length excluding silence intervals. Each song is accompanied by MIDI and text annotations. We used a rule-based Korean grapheme-to-phoneme method \cite{cho2017kog2p} to preprocess raw text. 45 songs were used for training, one song for validation, and four songs for testing. To generate ground-truth mel-spectrograms, we downsampled the recordings to 16 kHz, applied a pre-emphasis factor of 0.97, then used a filter length of 1024, hop size of 200, window length of 800, and $D_\text{mel}$ of 120 for STFT.

\subsection{Setup}

We used 16 Mixer blocks as the backbone of MLP Singer. The text embedding dimension $D_c$ was set to 256; $D_p$, 32. $D_p$ was kept deliberately small since pitch embeddings only have to express a total of 25 notes including a silence token. The maximum sequence length $L$ was set to 200 frames, which translates to 2.5 seconds. If $L$ is excessively large, the model would unnecessarily attempt to learn noisy correlations between frames that are too distant, whereas the converse would limit the receptive field of the model. Dropout probability was set to 0.5 for all layers. We expect the model to perform better with more exhaustive hyperparameter search. 

MLP Singer was trained on a single NVIDIA RTX 6000 GPU with a batch size of 384, using the Adam optimizer \cite{kingma2017adam} with $\beta_1 = 0.9$, $\beta_2 = 0.999$. The learning rate was set to 0.001 and adjusted with linear warmup followed by linear decay. 

\subsection{Evaluation}

We compared MLP Singer with BEGANSing, an autoregressive conditional GAN-based model proposed by Choi et al. \cite{choi2020korean}. While we also considered including Adversarial SVS \cite{lee2019adversarially}, we were unable to reproduce the results of the paper, nor were there open source resources. The official implementation of BEGANSing was publicly available. 

For the vocoder, we used a variant of HiFi-GAN \cite{kong2020hifi}, which was fine-tuned on CSD to account for distributional differences between mel-spectrograms produced from speech and singing voice recordings. Since BEGANSing outputs magnitude spectrograms, we applied a mel-basis projection to convert synthesized outputs to mel-spectrograms. We evaluated MLP Singer under two configurations: naive batching and overlapped batch segmentation as delineated in Section \ref{sec:batch}, with $w$ set to 30. Lastly, ground-truth recordings and their vocoder reconstructions were included as points of reference to estimate vocoder-induced bottleneck. We evaluated the SVS systems on both audio quality and inference speed. 

\subsubsection{Audio Quality}

8 audio segments were randomly selected from the test set. We conducted a mean opinion score (MOS) evaluation with 10 participants.\footnote{Audio samples available at \url{https://mlpsinger.github.io}.} Each sample was rated from a scale of 1 to 5, in 0.5 increments. MLP Singer and overlapped batch segmentation are abbreviated as ``MLPS" and ``OB", respectively.

\begin{table}[th]
  \centering
  \begin{tabular}{ c|cc }
    \toprule
    Model & Params. (M) $\downarrow$ & MOS $\uparrow$ \\
    \midrule
    BEGANSing \cite{choi2020korean} & $42$ & $2.325 \pm 0.144$ \\
    MLPS (Ours) & $\mathbf{8}$ & $2.875 \pm 0.149$ \\
    MLPS + OB (Ours) & $\mathbf{8}$ & $\mathbf{3.169 \pm 0.153}$ \\
    \midrule
    GT & - & $4.269 \pm 0.105$ \\
    Reconst. & - & $3.575 \pm 0.153$ \\
    \bottomrule
  \end{tabular}
  \label{tab:eval}
  \caption{Parameter count and MOS}
\end{table}

Overall, MLP Singer received higher ratings than the baseline model. The result also suggests that overlapped batch segmentation positively impacts perceived audio quality. A paired sample \textit{t}-test on the MOS of MLP Singer with and without overlapped batching indicated a statistically significant difference at a 95\% confidence interval given a \textit{p}-value of $3.525 \times 10^{-5}$, suggesting that overlapped batch segmentation helped produce better results.

Vocoder reconstructions received considerably lower scores compared to ground-truth recordings. We suspect that this was due to noisy artifacts that the vocoder produced at high frequency intervals. Given the non-negligible vocoder bottleneck, the MOS for ground-truth reconstructions can be viewed as the upper bound for an SVS system. 

\subsubsection{Inference Speed}

Inference speed tests were conducted on an Intel Core i9 CPU and NVIDIA RTX 6000 GPU. We measured the time it took for each model to produce a given number of mel-spectrogram frames, including data operations such as tensor reshaping and slicing. Each measurement was taken 20 times to produce reliable estimates.

\begin{figure}
\centering
\includegraphics[scale=0.6]{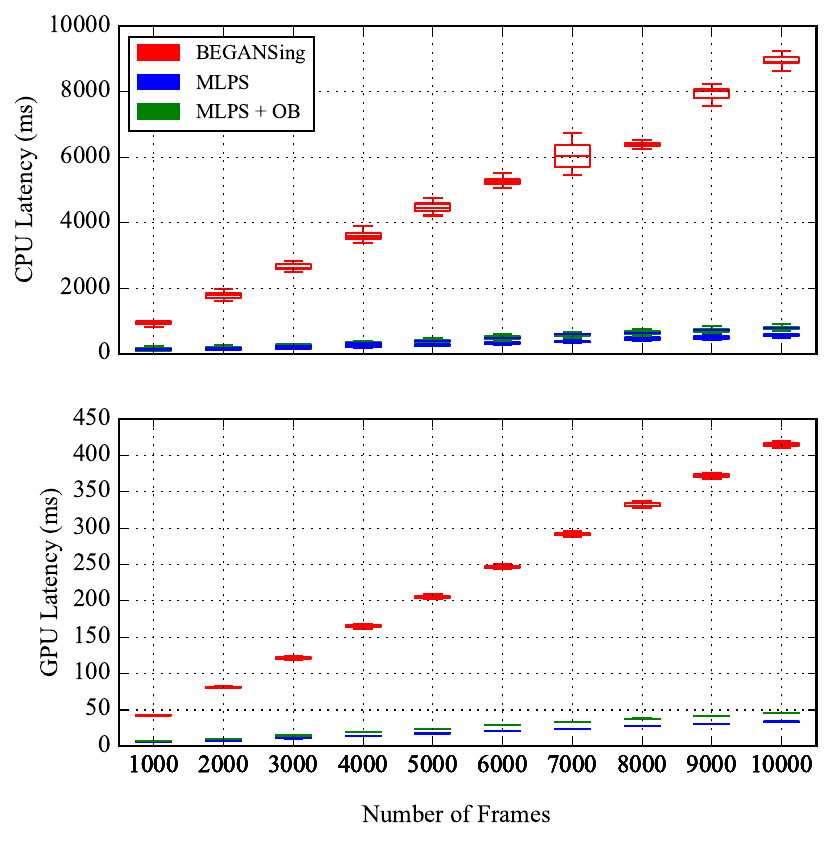}
\caption{Inference latency comparison on CPU and GPU}\label{fig:speed}
\end{figure}

As shown in Figure \ref{fig:speed}, MLP Singer consistently outperforms the autoregressive baseline on both GPU and CPU. As the number of predicted mel-spectrogram frames increases, BEGANSing's inference latency grows conspicuously, whereas MLP Singer slopes only slightly, even with the added overhead caused by overlapped batch segmentation. We also found a substantial gap in the real-time factor (RTF), which was computed by dividing the total duration of synthesized speech by inference latency. The average RTF of MLP Singer on CPU and GPU were 203 and 3401 (145 and 2514 with overlapped batch segmentation), whereas BEGANSing achieved approximately 14 and 301, respectively. MLP Singer's CPU latency is thus comparable to that of BEGANSing on GPU runtime, which demonstrates the scalability and efficiency of the proposed system.

\section{Discussion}

\subsection{Shift Invariance}

Initially, the Token Mixer appears inappropriate for processing sequential data since features in each time step get multiplied with a different set of weights. This could be undesirable since we expect the model to generate roughly identical acoustic features regardless of where a particular pitch-syllable combination appears in the input chunk.

To further investigate this intuition, we conducted an ablation study in which we removed Token Mixers from MLP Singer and constructed a model purely composed of Channel Mixers. This ablated model is essentially a stack of fully-connected layers with a receptive field that is limited to a single time step. To ensure that model performance is not affected by differences in parameter count, we configured the ablated model to have 24 layers, which amounts to 8M parameters. Despite this adjustment, the model suffered a noticeable drop in performance. We quantitatively evaluated this degradation by measuring the L1 mel-spectrogram loss on the holdout set. We also computed mel-cepstral distortion (MCD) to introduce an additional metric unused during training.

\begin{table}[th]
  \centering
  \label{tab:ablation}
  \begin{tabular}{ c|ccc }
    \toprule
    Model & L1 Loss $\downarrow$ & MCD $\downarrow$ \\
    \midrule
    MLP Singer & $\mathbf{0.0725}$ & $\mathbf{1.87}$ \\
    Channel Mixer-only & $0.0837$ & $2.26$ \\
    \bottomrule
  \end{tabular}
  \caption{Token Mixer ablation results}
\end{table}

The relatively high L1 loss and MCD values indicate that the ablated model failed to learn quality transformations that approximate ground-truth spectrograms. Given that Channel Mixers and Token Mixers operate on orthogonal dimensions, the result suggests that transformations on one axis cannot substitute the effect of that on the transposed dimension.

\begin{figure}
  \centering
  \includegraphics[scale=0.4]{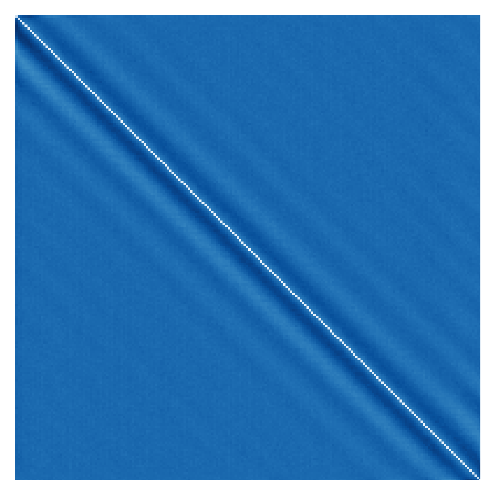}
  \includegraphics[scale=0.4]{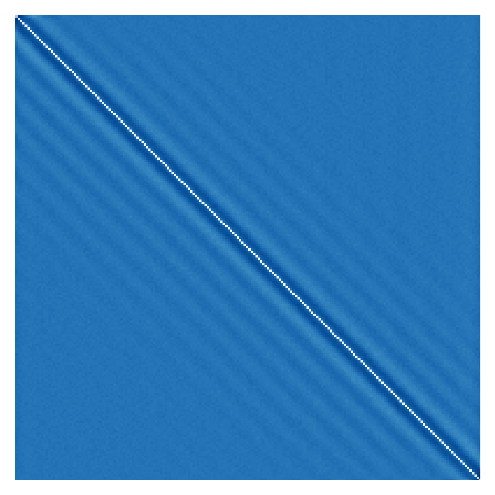}
  \caption{Heatmaps of learned Toeplitz-like transformations}
  \label{fig:weights}
\end{figure}

By examining the trained weights of the model, we found that Token Mixers learned Toeplitz-like transformations that reflect shift invariance. Since a Token Mixer is composed of two fully-connected layers, we fed an identity matrix $\mathbb{I}_L$ and plotted the output to visualize their aggregate effect. Figure \ref{fig:weights} displays the output produced by 8th and 11th Token Mixers. Previous work that explored similar MLP-based models \cite{liu2021pay} found that masked language modeling \cite{devlin2019bert} encouraged models to learn shift invariance via Toeplitz transformations, since any offset of the input should not affect the task of predicting a masked token. Similarly in SVS, moving the text and MIDI inputs by some number of frames should not result in drastically different mel-spectrograms. Despite the lack of an explicit inductive bias, MLP Singer becomes robust to temporal shifts by learning Toeplitz-like matrices, which are characterized by their constancy along the descending diagonal. Thus, regardless of where a particular feature appears in an input segment, it is multiplied by a similar set of weights and undergoes a shift-invariant transformation.

\subsection{Frame Congruence}

\begin{figure}
  \centering
  \includegraphics[scale=0.6]{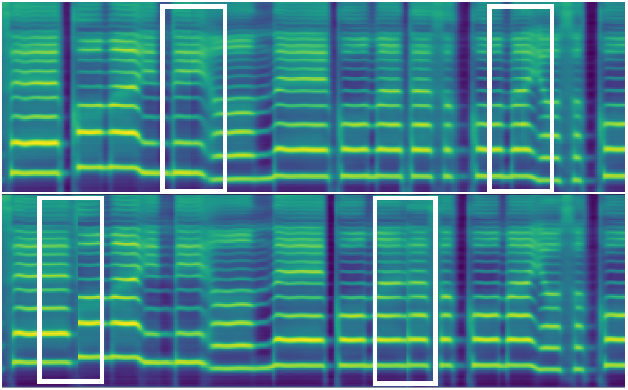}
  \caption{Segment boundaries in synthesized mel-spectrograms (without overlapped batch segmentation)}
  \label{fig:mel}
\end{figure}

Non-autoregressive models are known to produce poorly connected spectrogram frames since each chunk is synthesized independently of one another \cite{blaauw2017npss,choi2020korean}. Generally, we found that MLP Singer is capable of generating continuous frames on intervals with smooth harmonic patterns. Since the model has learned shift-invariant transformations, it can produce roughly similar outputs regardless of how the input is chunked. The top row of Figure \ref{fig:mel} illustrates an instance of felicitous generation; the bottom row displays a counter-example that contains jagged artifacts near segment edges. The two outputs only differ by how the input was chunked. Segment boundaries are indicated as boxes.

\begin{figure}
  \centering
  \includegraphics[scale=0.25]{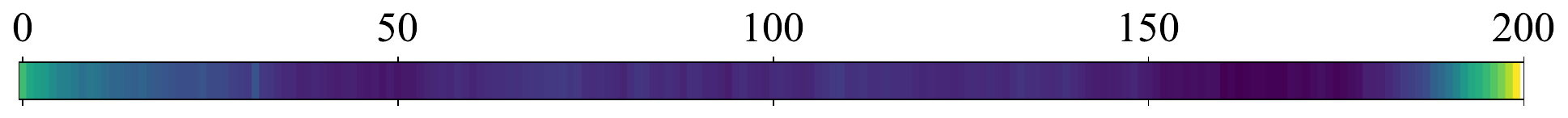}
  \caption{Average per-location mel-spectrogram loss}
  \label{fig:heatmap}
\end{figure}

Empirically, we found that visible junctures occur when segment boundaries coincide with transitions, such as pitch changes, rapid modulations, or movement from voiced to unvoiced intervals. We speculate that this is due to the limitation of Toeplitz transformations. Given a Toeplitz-like matrix with $2r$ diagonally constant components, shift invariance does not apply to the first and last $r$ frames in the input segment. In other words, embeddings in the first few time steps are only mixed with features in the future, whereas the last few representations are only mixed with those from the past. In contrast, the Token Mixer can produce rich representations for frames that are located in the the middle of the chunk between indices $[r, L - r]$, since it has sufficient bidirectional access to features $\pm r$ frames away from the current time step. This analysis is coherent with the observed average per-location mel-spectrogram loss on the test set, which is highest at each ends of the model's receptive field as shown in Figure \ref{fig:heatmap}. The overlapping batch segmentation method mitigates this problem by enabling the model to look $w$ steps into the past and future at segment boundaries via overlaps. The theoretically optimal value of $w$ would then equal $r$.

Transitions and segment boundaries rarely coincide by chance, which is why MLP Singer can reliably produce felicitous outputs. Experiments also show that overlapping batch segmentation leads to better perceived audio quality. Nonetheless, improved batching is an inference time sampling technique rather than a fundamental architectural improvement. We leave this area for investigation in future work. 

\section{Conclusion}

We present MLP Singer, an all-MLP parallel Korean singing voice synthesis system. MLP Singer demonstrates competitive edge over an existing autoregressive GAN baseline both in terms of synthesis quality and inference latency, achieving order of magnitude faster real-time factors on both CPU and GPU by leveraging parallel generation. We analyze the effectiveness of MLPs in terms of their ability to learn shift invariance as Toeplitz-like transformations. We also propose a batch segmentation method that uses overlapping windows to ensure the model sufficient bidirectional receptive field, thus preventing it from generating discontinuous frame artifacts that typically affect non-autoregressive systems. 

\section{Acknowledgement}

We express gratitude to Juhan Nam of KAIST Music and Audio Computing Lab for valuable discussions and feedback.

\bibliographystyle{IEEEbib}
\setlength{\bibsep}{0pt}
\renewcommand{\bibsection}{\section{References}}
\bibliography{refs}

\end{document}